\documentclass[review]{elsarticle}

\usepackage{amssymb}
\usepackage{graphicx}
\usepackage{times}
\usepackage{amsmath}
\usepackage{amsfonts}
\usepackage{textcomp}
\usepackage{multirow}
\usepackage{array}
\usepackage{url}
\usepackage[usenames,dvipsnames]{color}
\usepackage{float}
\usepackage{subfigure}
\usepackage{color}
\usepackage{pgfplots}
\usepackage{fancyhdr}
\usepackage{siunitx} 
\usepackage{booktabs}

\journal{International Journal of Solids and Structures}

\begin{document}

\begin{frontmatter}

\title{Bending of Biomimetic Scale Covered Beams Under Discrete Non-periodic Engagement}

\author[mymainaddress]{Hessein Ali}
\author[mymainaddress]{Hossein Ebrahimi}
\author[mymainaddress]{Ranajay Ghosh\corref{mycorrespondingauthor}}
\cortext[mycorrespondingauthor]{Corresponding author}
\ead{ranajay.ghosh@ucf.edu}

\address[mymainaddress]{Department of Mechanical and Aerospace Engineering, University of Central Florida, Orlando FL, 32816}

\begin{abstract}
Covering elastic substrates with stiff biomimetic scales significantly alters the bending behavior via scales engagement. This engagement is the dominant source of nonlinearity in small deflection regime. As deformation proceeds, an initially linear bending response gives way to progressive stiffening and thereafter a geometrically dictated `locked' configuration. However, investigation of this system has been carried out until date using assumption of periodic engagement even after scales contact. This is true only under the most ideal loading conditions or if the scales are extremely dense akin to a continuum assumption on the scales. However, this is not true for a practical system where scales are more discrete and where loading can alter periodicity of engagement. We address this nonlinear problem for the first time in small deflection and rotation regime. Our combined modeling and numerical analysis show that relaxing periodicity better represents the geometry of discrete scales engagement and mechanics of the beam under general loading conditions and allows us to revisit the nonlinear behavior. We report significant differences from predictions of periodic models in terms of predicting the behavior of scales after engagement. These include the difference in the angular displacement of scales, normal force magnitudes along the length, moment curvature relationship as well as a distinct nature of the locking behavior. Therefore, non-periodicity is an important yet unexplored feature of this problem, which leads to insights, absent in previous investigations. This opens way for developing the structure-property-architecture framework for design and optimization of these topologically leveraged solids.
\end{abstract}


\end{frontmatter}


\emph{}\section{Introduction}

\begin{figure}[htbp]
\centering
\subfigure[]{%
\includegraphics[scale = .046]{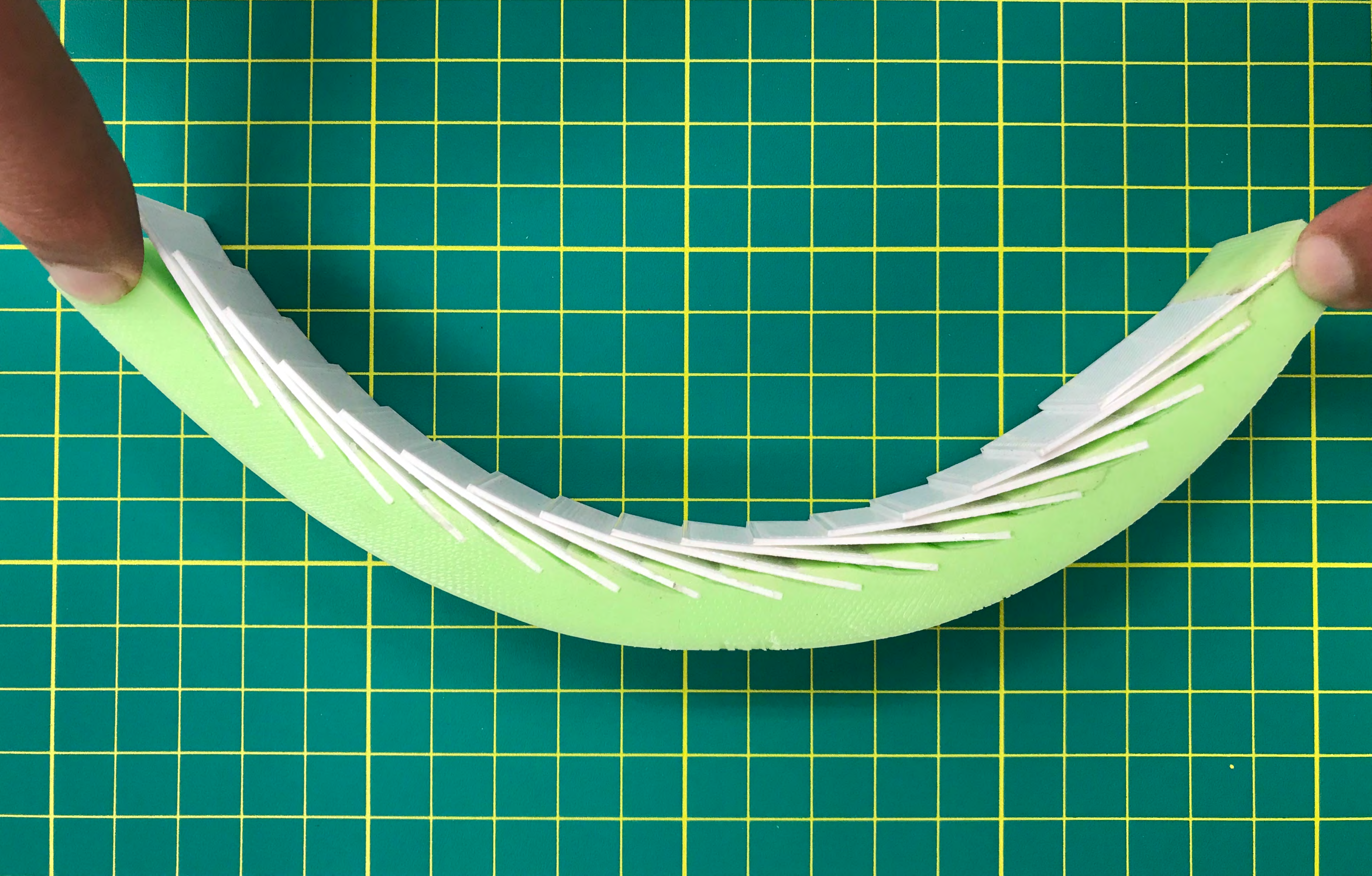}
\label{fig1A}}
\quad
\subfigure[]{%
\includegraphics[scale=0.038]{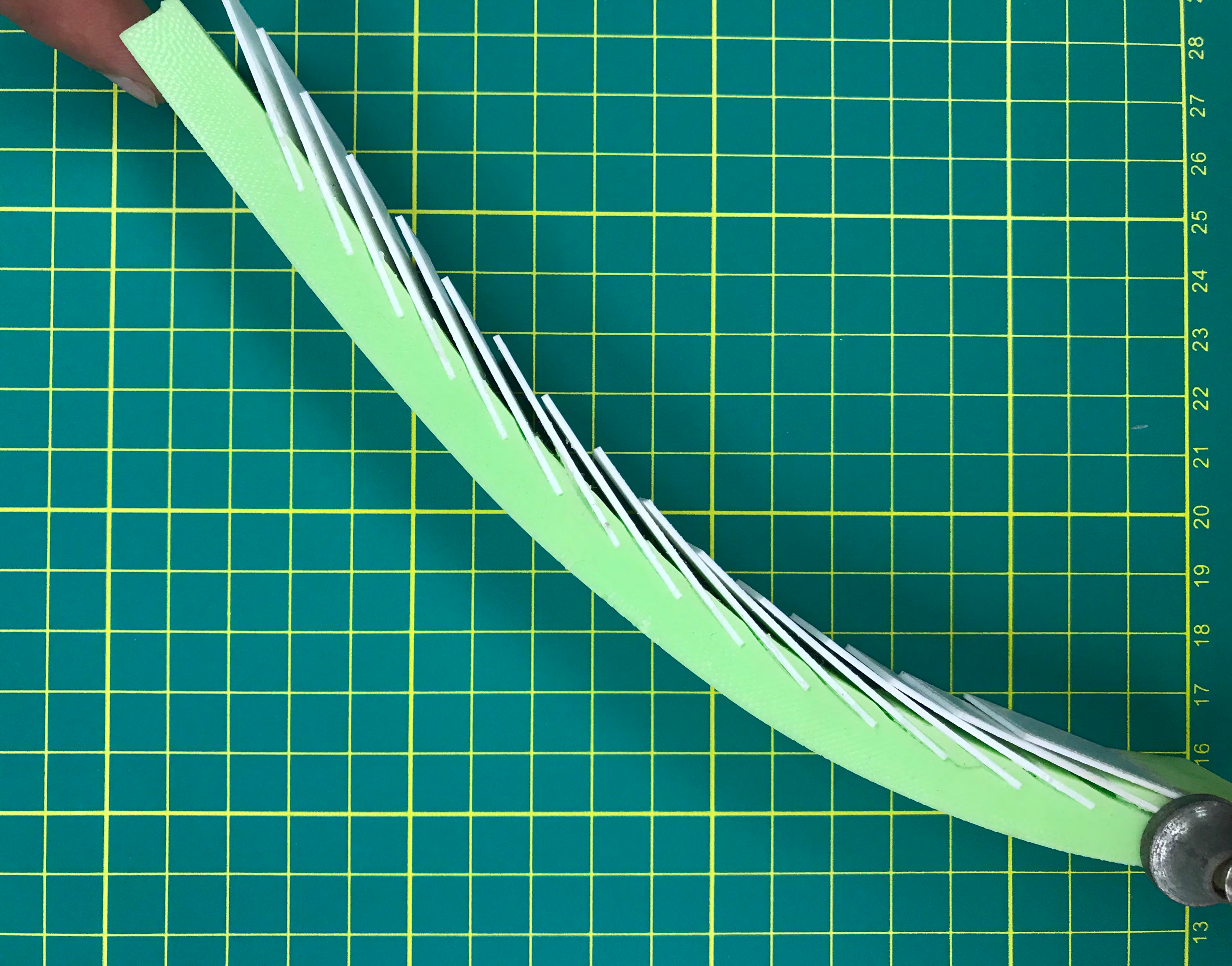}
\label{fig1B}}
\caption{(a) A manual illustration of periodic engagement of scales as the underlying structure bends uniformly. (b) An illustration of non-periodic engagement of scales through the example of a cantilever scaly beam. The substrate and scales were printed using Polylactic Acid (PLA) and Vinylpolysiloxane (VPS), respectively. The dimensions of the fabricated substrate are $200$ mm (length) x $25$ mm (width) x $5$ mm (height) while scale dimensions are $35$ mm x $25$ mm x $1$ mm with inclination angle of $10^{\circ}$. The spacing between scales is $10$ mm.}
\label{fig1} 
\end{figure}

Biological structures have inspired synthetic materials with unparalleled performances such as ultra-lightweight design~\cite{lightweight1,lightweight2,lightweight3,lightweight4}, tunable elasticity~\cite{tune1,tune2,tune3,tune4,tune5}, and negative poisson's ratio~\cite{neqposs1,neqposs2,neqposs3,neqposs4,neqposs5}. Among biological structures, scales had appeared in the earliest stages of evolution of complex multicellular life~\cite{bruet2008materials} and continued their existence in spite of millions of years of evolutionary pressures. This has made scales a naturally high performance material with hybrid and multiscale response to various loads~\cite{perform1,perform2,perform3,perform4,perform5,perform6,perform7,nonlinear1}. For instance, scale covered organisms have inspired dermal armors fabricated using a soft substrate with plate-like ceramics embedded on the top layer~\cite{armor1}. This design showed that overlapping of scales provides flexibility, damage tolerance, and more importantly resistance to puncture. Similarly, armadillo scales have also been used as a source of inspiration for designing flexible armor fabricated using hexagonal glass plates placed on an elastomer substrate~\cite{armor2}. This type of synthetic armors also yielded a good resistance to puncture as well as flexibility. In addition, the development of flexible armor has also been implemented on fabrics~\cite{armor1,armor3}. However, in addition to material response of the scales themselves, the scales serve as topological modifications to the underlying substrate. This `structural' as opposed to the purely material aspect of scales reveals an entirely different regime of response encompassing interesting nonlinear behavior. In this case, typically scales are attached to a low dimensional flexible substrate such as a beam or a plate. In such cases, in contrast to armor like `local' loading, scale arrangement influences global deformation behavior such as bending as the biomimetic scale beam shown in Fig.~\ref{fig1A}. For such scaly substrates, mechanical behavior depends critically on the kinematics of scale sliding. 

In this context, particularly, scaly structures subjected to a pure bending moment have been intensively investigated due to their practical and theoretical importance in isolating kinematics and developing moment curvature relationships. For instance, in one of the earliest studies, the mechanism of deformation of a fish scale structure (with the assumption of deformable scales) was investigated where the authors demonstrated the strain-stiffening response in the structure~\cite{mechanics1}. Further work on deformable scales followed investigated stretch and buckling response of teleost fish structures~\cite{perform4}. To address the mechanics of two-dimensional scaly composite shells, a computational approach was proposed~\cite{2dmechanics} to establish the relationship between structure and the mechanical response. The authors studied the structure under both bending and twisting types of loading. These studies clearly showed that stiffer scales at a low angle are desirable for maximum performance. Taking this route and simplifying such a high contrast system (stiff scales and soft substrates)  with rigid scales helps isolate the role of scale kinematics on the mechanical nonlinearity. This simplified assumption leads to closed form analytical relationships connecting the kinematics to the mechanics. In this context, the kinematics and mechanics of a one-dimensional scaly beam, assuming rigid scales, have been addressed~\cite{mechanics2}. In this work, the authors assumed frictionless self-contact between scales. Their results revealed the existence of a three different regimes of mechanical response - linear, non-linear, and locking phase. The effect of friction in sliding kinematics of scales has then been further studied in~\cite{mechanics3}. The study revealed that friction does not alter the overall nature of behavior although it advanced the locking envelopes further. Further follow up studies which outlined the envelopes of validitity of the analytical models for rigid scale system were also carried out using extensive finite element (FE) analysis~\cite{mechanics4}. Furthermore, composite architecture with scales only embedded on the top layer of a soft substrate (imitating elasmoid fish scales) have been presented to account for the deformation mechanism due to compressive loading~\cite{compositefish1,compositefish2}.  In their work, the authors found that volume fraction of the embedded plate like scales has a prime role in changing the stiffness of an elastomer structure.

These prior investigations underscore the growing importance of using scales as topological additives on substrates. In order to fully develop the structure-property-architecture paradigm for this class of hybrid materials, models are of critical importance. This is because they do not only reveal and quantify the mechanism of nonlinear behavior but also indispensable for design and optimization of the architecture.  Therefore, it is imperative that models accurately reflect salient aspects of the system. Thus far, all models have relied on the assumption of preserving periodicity throughout scales engagement. This assumption allows the isolation of a fundamental representative volume element (RVE), after which periodic boundary conditions are applied and a global derivative is affected to obtain the mechanical behavior~\cite{perform4,mechanics1,mechanics2,mechanics3}. However, in any realistic structural application such post-engagement periodicity is seldom observed either at a global or local level beyond the simplest of the loading cases such as pure bending (see Fig.~\ref{fig1A}). Periodicity of engagement can be broken by simply applying different boundary and loading conditions.  For instance, a cantilevered beam would not exhibit periodicity associated with pure bending. This is approximately shown in the contrasting geometries post engagement between Fig.~\ref{fig1A} and ~\ref{fig1B}. In fact, the density of scales needed to maintain even local periodicity for such cases is considerable and typically not observed in real systems which have discrete scales distribution. More importantly, an enormously dense scale system begins to mask the tunable nonlinearity specific to scale sliding due to the material constriction effect between the scales ~\cite{mechanics2,compositefish1}. Last but not the least, even for global periodicity, the number of scales in real structures are often not sufficient to justify a continuous distribution. 

In spite of these known limitations, existing models still rely on periodic frameworks which cannot be directly applied or even extended to the non-periodic cases such as the case of a cantilever beam illustrated in Fig.~\ref{fig1B}. Therefore, it is imperative that investigation be based on more accurate models which could address the lack of periodicity and discrete nature of the scales. This work presents a more general theory of stiff scale covered elastic substrate to establish the kinematics and mechanics of a one-dimensional scaly beam using scale-by-scale interaction approach obviating the need for global or local periodicity. The theory is first applied to structures that undergo a uniform bending which are compared with results in literature~\cite{mechanics2}. The model is then validated using FE-based numerical studies to show the accuracy of our theory. Kinematics and mechanics of non-uniform bending structures will also be presented for the cases of simply supported and cantilever beams.  The analytical results show a perfect match with FE results which prove that no other mechanical assumptions are needed to explain previous discrepancies.  

\section{Materials and Methods}
 
\begin{figure}[htbp]
\centering
\includegraphics[scale = 0.4]{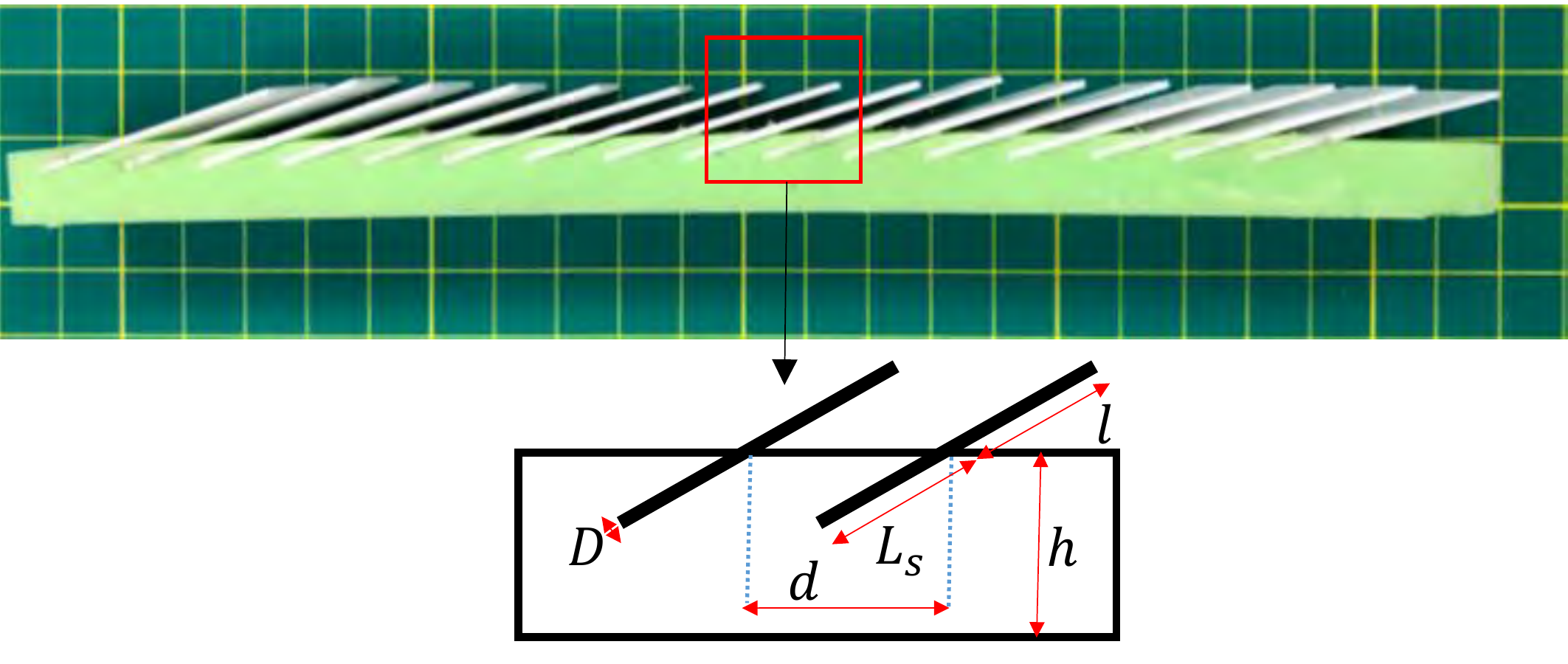}
\caption{ The reference configuration of scaly biomimetic system, and a schematic diagram of two neighboring scales. The sample has the same dimensions as the one illustrated in Fig.~\ref{fig1}}.
\label{figG}
\end{figure}

\textbf{Geometry:} The geometry of the system in the reference configuration is illustrated in Fig.~\ref{figG}. A periodic arrangement in the reference configuration is apparent. The underlying substrate is assumed to be a uniform beam of length $L$. The length of the scale is assumed to be $l_s=l+L_s$ where $l$ is the exposed part of the scale and $L_s$ is the embedded part. The thickness of the scale is considered to be $D$ and the beam thickness is $h$. It is further assumed that $D \ll L_s$ and $h \gg L_s$, an assumption commonly made indicating scales thin are confined to the top of the substrate. We denote the ratio of scale length to separation as $\eta=l/d$ where $d$ is the distance between the scales. The scales start with an initial scale angle $\theta_0$ measured with respect to the beam centerline and rotates to an angle $\theta$ as the engagement proceeds.  

\textbf{Materials:} A typical scaly biomimetic system features scales which are much stiffer than the underlying substrate. This study targets a system which could be comprised of a silicone based substrate of modulus $E= 1.5$ MPa and poisson's ratio $\nu=0.23$ and PLA plastic for scales with $E= 2.86$ GPa. Clearly the moduli are widely divergent for these materials which allows for treating the scales as rigid as long as locking conditions are not realized~\cite{perform4,mechanics4}. The strains are assumed to remain small and the beam can be approximated by the Euler-Bernoulli assumptions. A further un-stretchable constraint on the beam is imposed.
 
\textbf{Kinematics:} The periodicity after contact is a typically strong constraint and will be readily violated via boundary and loading conditions for a practical system.  An example of this is the case of non-uniform bending such as cantilever or distributed loading. Local periodicity, however, could be maintained for very high density of scales but  that would transition this system to a more composite and coating type systems dictated by material constrictions ~\cite{compositefish1}. 

\begin{figure}[htbp]
\centering
\subfigure[]{%
\includegraphics[scale = .73]{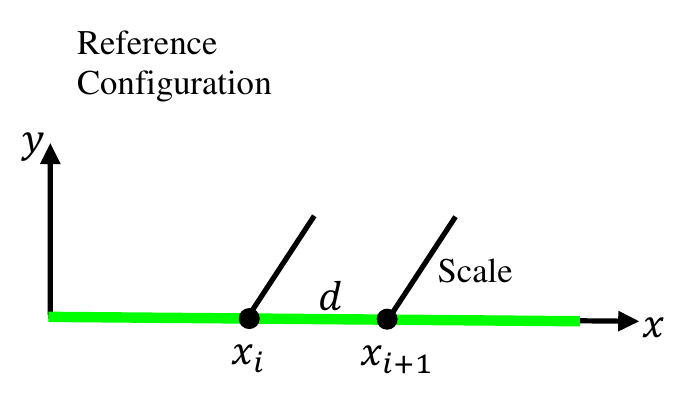}
\label{fig2A}}
\quad
\subfigure[]{%
\includegraphics[scale =0.67]{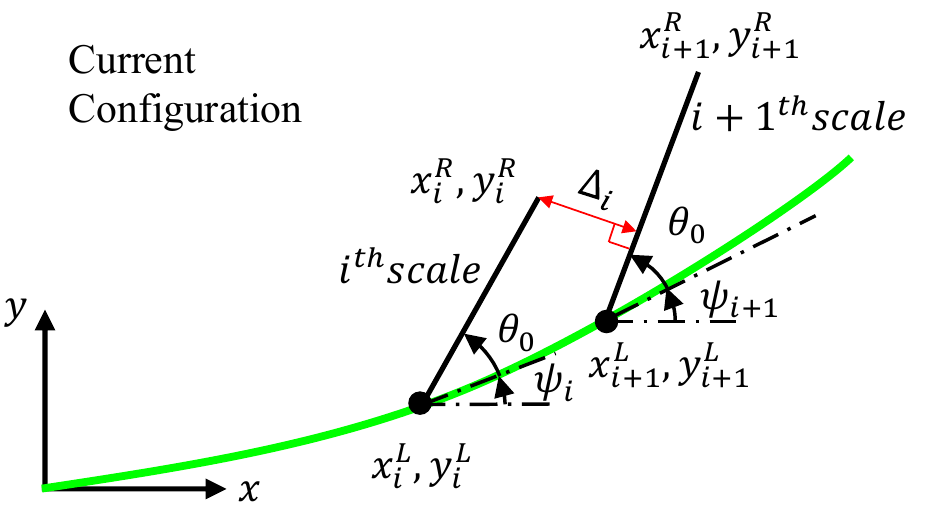}
\label{fig2B}}
\caption{(a) A geometry of a beam with scales at the initial configuration. (b) A configuration of the deformed beam and scale geometry before engagement.}
\label{fig:figure}
\end{figure}

In order to address the breakdown of periodicity, a scale-by-scale discrete approach is introduced in this work. It is assumed that in the reference configuration, the position of the $i^{th}$ scale on the substrate is given by  $x_i$. A general material point on the substrate in the reference flat state is denoted by $x$. This is shown in Fig.~\ref{fig2A}. The current configuration of the scale is quantified by the coordinates $x_i^L,x_i^R$ which are the left and right ends of the scale as shown in Fig.~\ref{fig2B}. In the case of pure bending, the typical measure of deformation is the curvature. However, for more general loading case an alternative way to devise deformation is presented in this paper using a shape function $f(x)$ and its normalized amplitude  $\gamma$ which determines the extent of load. Therefore, in the current configuration, the material point now occupies a vertical position $y(x)=\gamma f(x)$. In practice, $\gamma$ is a unit less constant which depends on the load, beam geometry and substrate material.  In pure bending, moment causes a substrate to deform into an arc. In small deflection, this arc will follow the form  $y(x)= \kappa (1/2 x^2-Lx/2)$ with the instantaneous curvature $\kappa = M/EI$ where $L$ is the length of the beam, $M$ is the bending moment, and $EI$ is the flexural rigidity of the beam~\cite{shigleybook}. We non-dimensionalize the curvature with the beam thickness to get $\gamma=\kappa h$. On the other hand, the deflection of a simply supported beam of flexural rigidity $EI$ and uniform loading $w_0$ has the form $y(x)={w_0 \over 24EI}(2Lx^3-x^4-L^3 x)$. In this case,  $\gamma={w_0h^3\over24EI}$ . Finally a cantilever beam with point load $p_0$  at the tip deforms according to the function $y(x)={p_0 \over 6EI}(x^3-3Lx^2)$ which makes $\gamma ={p_0h^2\over 6EI}$~\cite{shigleybook}. 

With the assumption of unstretchability, a scale level geometry, shown in Fig.~\ref{fig2B}, emerges before engagement commences. From this geometry, we can write for any scale, before engagement:
\begin{align}
x_i^R & = x_i^L + l\cos({\theta_0 +\psi_i}), \notag \\
 y_i^R & = y_i^L + l\sin({\theta_0 +\psi_i}) , \label{eq1} \\
\tan({\psi_i}) & = \gamma f'(x_i^L), \text{and}\ x_i^L \equiv{x_i}. \notag
\end{align}

Where $\theta_0$ is the initial inclination angle of the $i^{th}$ scale and $\psi_i$ is the inclination angle of the beam at the base of the $i^{th}$ scale. This geometry will undergo further change as engagement proceeds. The scales engagement can be tracked using the distance parameter $\Delta_i$ of the right extremity of the scale to the subsequent scale as shown in Fig.~\ref{fig2B}. This distance parameter can be written as~\cite{thomascalculusbook}:
\begin{equation}
\Delta_i= {1 \over l}((y_{i + 1}^L - y_{i + 1}^R)(x_{i + 1}^L - x_i^R) - (x_{i + 1}^L - x_{i + 1}^R)(y_{i + 1}^L - y_i^R)),i = 1,..,{N_s} - 1
\label{eq2}
\end{equation}
Where $N_s$ is the total number of scales. As $\Delta_i$ becomes zero, engagement condition is met. 

\begin{figure}[htbp]
\centering
\subfigure[]{%
\includegraphics[scale = 0.75]{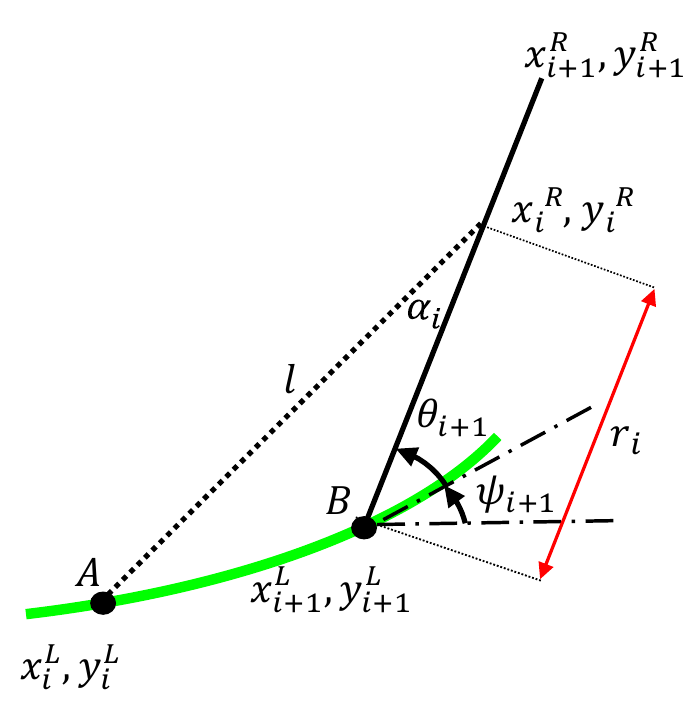}
\label{fig3A}}
\quad
\subfigure[]{%
\includegraphics[scale = 0.75]{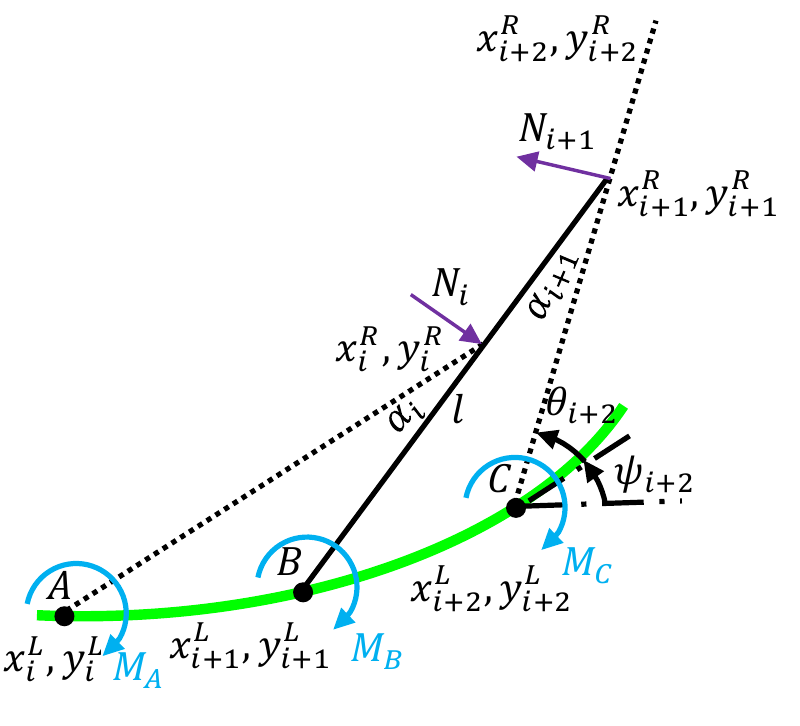}
\label{fig3B}}

\caption{(a) A deformed beam and scale geometry at a general point of engagement.  (b) A FBD of an individual scale.The dotted-adjacent scales are added for clarification.}
\label{fig:figure}
\end{figure}

To illustrate the effect of geometry change after engagement,  two sequential scales  $i$ and $i+1$ at a general point of engagement is taken. This is shown in Fig.~\ref{fig3A}. At this point, scale $i$ is engaged with scale $i+1$. After engagement, the geometry is constrained. The kinematics is governed by Eq.~\eqref{eq1} with $\theta_0$ replaced with $\theta_i$.  Moreover, $\theta_i$ and $\theta_{i+1}$ are both unknown, which makes the geometry statically indeterminate. To resolve this impasse, an additional constraining condition utilizing the normal reaction moment balance between scales after engagement would be required. The scale rotation is modeled (similar to previous work~\cite{mechanics1,mechanics2}) as a linear torsional spring which rotates about a fixed point. The spring constant $K_B$ is known to follow the analytical expression $K_B=C_B ED^2 (L_s/D)^n$ where $E$ is the modulus of elasticity of the substrate and $C_B$,$n$ are constants with values $0.66,1.75$, respectively~\cite{mechanics2}. However, using a new set of finite element (FE) simulations, $C_B$ was found to a more accurate value of 0.86 to specifically account for small initial inclination angles ${\theta_0} < {10^{\circ}}$. In the case that $i+1^{th}$ scale is itself not engaged to $i+2^{th}$, there are four unknowns which are $\theta_i$  , $\theta_{i+1}$ , $x_i^R$, and $y_i^R$. In order to obtain these unknowns, four constraining conditions would be required. These conditions are: the fixed length of the scale due to rigidity, the vanishing distance parameter due to contact, and the moment balance at the base of the $i^{th}$  and $i+1^{th}$ scale using the free body diagram illustrated in Fig.~\ref{fig3B}.  Thus, the following equations emerge:
\begin{equation}
\tan \left( {{\theta _i} + {\psi _i}} \right) = {{y_i^R - \gamma f\left( {{x_i}} \right)} \over {x_i^R - {x_i}}},
\label{eq3}
\end{equation}

\begin{equation}
\tan \left( {{\theta _{i + 1}} + {\psi _{i + 1}}} \right) = {{y_i^R - \gamma f\left( {{x_{i + 1}}} \right)} \over {x_i^R - {x_{i + 1}}}}, \text{and}
\label{eq4}
\end{equation}
\begin{equation}
{\left( {x_i^R - {x_i}} \right)^2} + {\left( {y_i^R - \gamma f({x_i})} \right)^2} = {l^2}.
\label{eq5}
\end{equation}

The constraining condition using the balance of the moment at the base of the scales is slightly more involved. For the case of engagement of only two scales, balancing the moment about points $A$ and  $B$, Fig.~\ref{fig3B} yields 
\begin{equation}
{N_i} = {{{K_B}\left( {{\theta _i} - {\theta _0}} \right)} \over {l\cos\left( {{\alpha _i}} \right)}} = {{{K_B}\left( {{\theta _{i + 1}} - {\theta _0}} \right)} \over {{r_i}}}.
\label{eq6}
\end{equation}
Where $\;{\alpha _i} = {\theta _{i + 1}} + {\psi _{i + 1}} - {\theta _i} - {\psi _i}$ and ${r_i} = \;\sqrt {{{\left( {x_i^R - {x_{i + 1}}} \right)}^2} + {{\left( {y_i^R - \gamma f({x_{i + 1}})} \right)}^2}} $ with $i = 1:N_s-1$. The fourth equation is now
\begin{equation}
\left( {{\theta _i} - {\theta _0}} \right){r_i} - l\cos\left( {{\alpha _i}} \right)\left( {{\theta _{i + 1}} - {\theta _0}} \right) = 0.
\label{eq7}
\end{equation}
 
The four highly nonlinear Eqs.~\eqref{eq3} through ~\eqref{eq5} and ~\eqref{eq7} must be solved numerically to obtain $\theta_i$  , $\theta_{i+1}$ , $x_i^R$, and $y_i^R$. Now extending this to a more general case of $N_e$ scales being engaged, the total number of unknowns would be: $N_e$ scale angles, $\theta_i$, $i=1:N_e$,  $2N_e-2$ coordinates of the right end of the scales $(x_i^R,y_i^R)$, $i=1:N_e-1$. Note that the coordinate requirements is reduced by one because the last scale undergoes no further engagements. This leads to a total of $3N_e-2$ unknowns. The total number of equations include the $N_e-1$ equations which correspond to constraint on the length of the scales and $2N_e-2$ which are based on the geometry of engagement of each scales excluding the last one. This yields a total of $3N_e-3$ equations. Finally, an additional equation is generated through the moment balance at the base at the last (far right) scale. Thus we now have a system of $3N_e-2$ unknowns and as many equations. Thus, balancing the moment about points $B$ and $C$,  Fig.~\ref{fig3B},  yields the following equation which can be utilized for finding the normal force between any two consecutive scales (except the case when $N_e=2$ or $i=1$, for which Eq.~\eqref{eq6} must be used): 
\begin{equation}
{N_{i + 1}} = {{{K_B}\left( {{\theta _{i + 1}} - {\theta _0}} \right) + {N_i}{r_i}} \over {l\cos\left( {{\alpha _{i + 1}}} \right)}} = {{{K_B}\left( {{\theta _{i + 2}} - {\theta _0}} \right)} \over {{r_{i + 1}}}}.
\label{eq8}
\end{equation}

Following the general procedure mentioned above, we can calculate the positions of all scales using a numerical solver such as available in commercial code MATLAB to maintain equilibrium at every step of deformation of the underlying substrate. Note that the angle of the right most scale will progressively decrease after engagement until it reaches an approximately zero angle. Accordingly, $\theta_{i+1}$ becomes known and the unknowns are only $x_i^R$ , $y_i^R$, and $\theta_i$. The structure now becomes statically determinate. Eqs.~\eqref{eq3} through ~\eqref{eq5} can then be simplified to uniquely determine the position of the $i^{th}$ scale. After simplification, Eqs.~\eqref{eq3} through ~\eqref{eq5}  yield the following quadratic equation:
\begin{equation}
(x_i^R-x_i)^2+\gamma f(x_{i+1})+{\gamma f'(x_{i+1})+\tan(\theta_{i+1})\over 1 - \gamma f'(x_{i+1})\tan(\theta_{i+1})}(x_i^R-x_i)-\gamma f(x_i)=l^2
\label{eq9}
\end{equation}

Equation.~\eqref{eq9} has only one unknown $x_i^R$ and gives the right $x-$coordinate of the $i^{th}$ scale. From this equation, one can obtain $y_i^R$ as
\begin{equation}
y_i^R   = \gamma f(x_{i+1})+{\gamma f'(x_{i+1})+\tan(\theta_{i+1})\over 1 - \gamma f'(x_{i+1})\tan(\theta_{i+1})}(x_i^R-x_i)
\label{eq10}
\end{equation}

and finally $\theta_i$ is calculated via Eq.~\eqref{eq3}. Note that Eqs.~\eqref{eq3},~\eqref{eq9}, and ~\eqref{eq10} are only utilized for finding the equilibrium configuration of the scales once the far right scale has reached an approximately zero inclination angle.

\textbf{Mechanics:} To better understand the mechanics of a scaly structure, one can envisage that the  bending mode is a combination of substrate bending and scales rotation. In other words, the structure stores energy during bending mode via the deflection of the beam and rotation of scales that is modeled as a linear torsional spring as described above ~\cite{mechanics1,mechanics2}. Thereafter, the mechanics of these structures is approached by employing the principle of minimization the total potential energy. We can write the total potential energy as $\Pi=\Omega_{beam}+ \Omega_{scales} H(-\Delta_i)-W$. Here $\Omega_{beam}$  is the strain energy of the underlying beam, $\Omega_{scales}$ is the strain energy due to the rotation of scales,  and $W$ is the work done by the applied load and $H$ is the Heaviside step function to track engagement. Since the deflection of the beam follows the form  $y=\gamma f(x)$, the energetic principle is equivalent to finding $\gamma$ that minimizes the potential energy through setting its first derivative to  zero. This leads to the following variational energetic equation ${d\Omega_{beam}\over d\gamma}+{d\Omega_{scales}\over d\gamma} H(-\Delta_i )=dW/d\gamma$. In general, the deflection will be characterized by the following two steps. First, we adopt $\gamma$ for the case of a virgin beam under appropriate loading conditions~\cite{shigleybook}. Once $\gamma$ is acquired, the second step becomes finding an equivalent load that balances the increase in the energy due to scales interaction.

For the case of uniform bending, the work done by an applied moment $M$ is $\mathop \smallint \limits_0^\kappa  Md\kappa '$ while the total energy stored in the system 
${1 \over 2}EI{\kappa ^2}L$ + $\mathop \sum \limits_{i = 1}^{{N_e}} {1 \over 2}{k_B}{\left( {\theta  - {\theta _0}} \right)^2} H(-\Delta_i).$  The moment-curvature relationship can be then expressed as: 
\begin{equation}
M = EI\kappa  + {1 \over L}\mathop \sum \limits_{i = 1}^{{N_e}} {k_Bh}\left( {\theta  - {\theta _0}} \right){{d\theta } \over {d\gamma }}\;H\left( { - {{\rm{\Delta }}_i}} \right).
\label{eq11}
\end{equation}

Here $d\theta/d\kappa$ is numerically evaluated for all the rotation angles of scales and their corresponding curvature. This relationship is equivalent to the one derived in earlier studies~\cite{mechanics1,mechanics2}.

Non-uniform bending is illustrated through the examples of simply supported and cantilevered beams. For the case of a simply supported beam subject to a uniform distributed load $w_0$, the work done can be written as $W = \;\mathop \smallint \limits_0^L {w_o}y\left( x \right)dx.$  The amplitude $\gamma$ is ${w_0h^3\over24EI}$ and therefore the deflection of the midpoint of the virgin beam can be expressed as $y_{mid}= {5w_0 L^4\over 384EI}$~\cite{shigleybook}. After engagement, the midpoint deflection will have the same formula. However, $w_0$ will be replaced with $w$ which is an equivalent load that provides the same midpoint deflection of a virgin beam including the effect of scales interaction. The equivalent load can be written as: 
\begin{equation}
w = {w_0} + {5h^3 \over {{L^5}}}\mathop \sum \limits_{i = 1}^{{N_e}} {k_B}\left( {\theta  - {\theta _0}} \right){{d\theta } \over {d\gamma }}.
\label{eq12}
\end{equation}

For simplicity, the midpoint deflection after the engagement of scales is written as $y_{mid}={5w L^4\over 384EI}$. 

Similarly, the work done in a cantilever beam due to a point load $p_0$ applied at the tip is $W=p_0 y(L)$. Therefore, the tip deflection is  $y_{tip}={p_0 L^3\over 3EI}$ while $\gamma={p_0h^2\over 6EI}$~\cite{shigleybook}. The interaction of scales will make $p_0$ increase in order to obtain an equivalent deflection in the case of having un-scaly substrate. This load is expressed in Eq.~\eqref{eq13}, and it is the alternative to $p_0$ to find the tip deflection after the interaction of scales begins.
\begin{equation}
p = {p_0} + {h^2 \over {2{L^3}}}\mathop \sum \limits_{i = 1}^{{N_e}} {k_B}\left( {\theta  - {\theta _0}} \right){{d\theta } \over {d\gamma }}.
\label{eq13}
\end{equation}

It is worth noting that the concept presented here can be applied to scaly structures with general types of loading and boundary conditions. Furthermore, to verify the kinematics and mechanics results of the three examples illustrated in this paper, finite element (FE) simulations using a commercially available code ABAQUS (Dassault systemes) were carried out. Several constraints were imposed on the models including rigid scales, surface-to-surface frictionless contact, and proper boundary conditions based on each example. Sufficient mesh density was used to ensure convergence in the results obtained. 
\section{Results and Discussion} 
\begin{figure}[H]
\centering
\subfigure[]{%
\includegraphics[scale = 0.6]{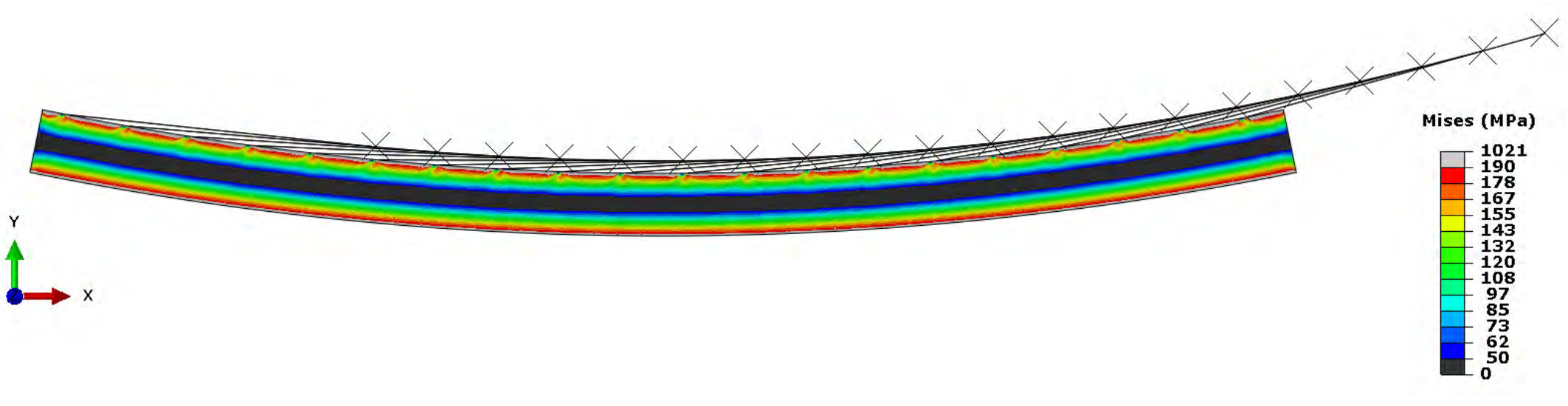}
\label{fig4A}}
\quad
\subfigure[]{%
\includegraphics[scale = 0.6]{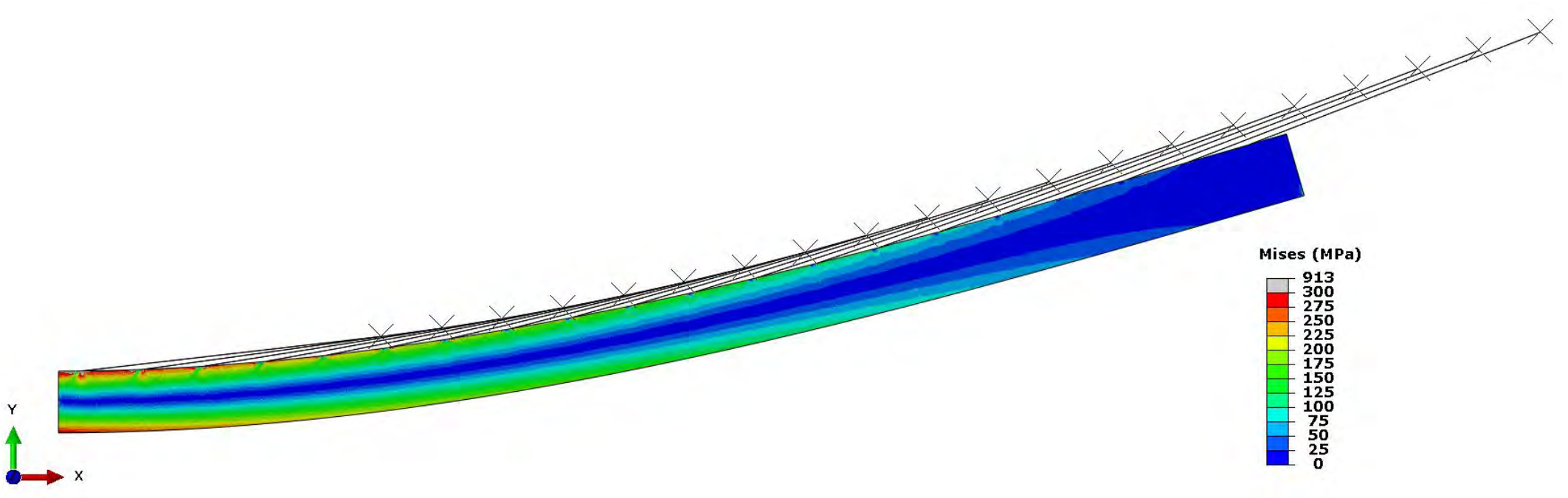}
\label{fig4B}}
\subfigure[]{%
\includegraphics[scale = 0.6]{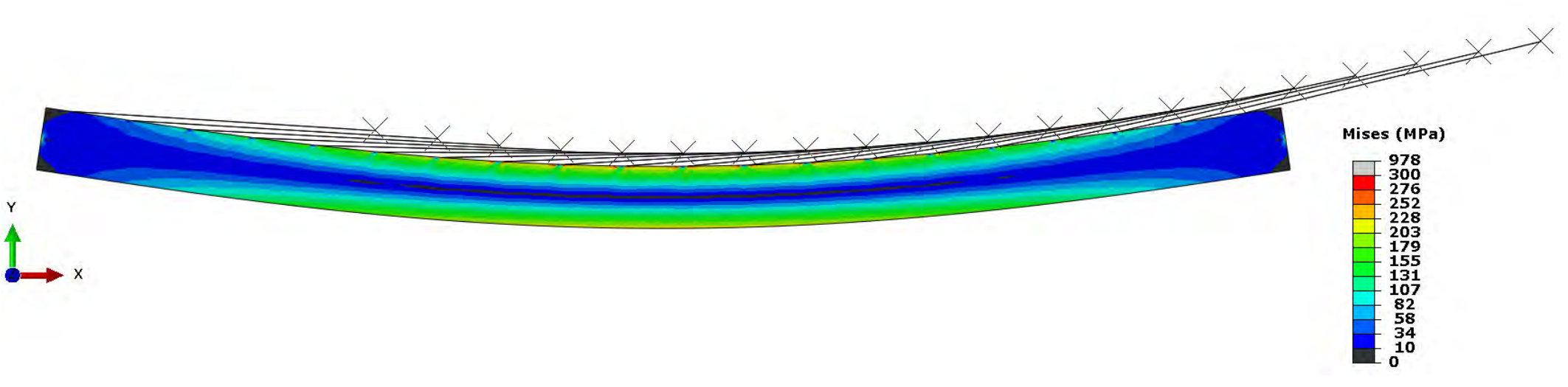}
\label{fig4C}}

\caption{(a) The initial engagement of scales, when the substrate uniformly deflects, with a contour plot of the vertical deflection of the beam at that instant. The dimensions of the substrate are $L$=1000 mm, $h$ = 50 mm while the scale dimensions are $l$=200 mm , $D$ = 0.05 mm, and $L_s$ = 7 mm. The substrate was assigned modulus of elasticity $E$=20 GPa and $\nu$ = 0.23. (b)  The breakdown of periodic engagement of scales of a cantilever scaly beam subject to a point load at the tip . (c)  An illustration of the lack of periodic engagement of scales via the example of a simply supported scaly beam loaded uniformly.}
\label{fig:figure}
\end{figure}

The regime where periodicity of scales engagement is preserved is first studied to compare with previous analysis. This is only observed when a scaly beam is uniformly bent, which is the case of applying a pure bending moment, and can be clearly seen in the von-Mises stress plots in the FE results shown in Fig.~\ref{fig4A}. The figure illustrates a uniform bending of a scaly beam consisting of $20$ scales in which the instant of engagement occurs at the same angle of curvature $\psi = \psi_i$. However, beyond this limited case of uniform bending, it is clear that periodic engagement of scales in no longer valid as illustrated by the non-uniformity of the von-Mises stress contours. This is the case for non-uniform bending of the underlying substrate such as a cantilevered scaly beam, Fig.~\ref{fig4B}, and the case of uniform loading on a simply supported scaly beam, Fig.~\ref{fig4C}.

\begin{figure}[t!]
\centering
\subfigure[]{%
\includegraphics[scale = 0.4]{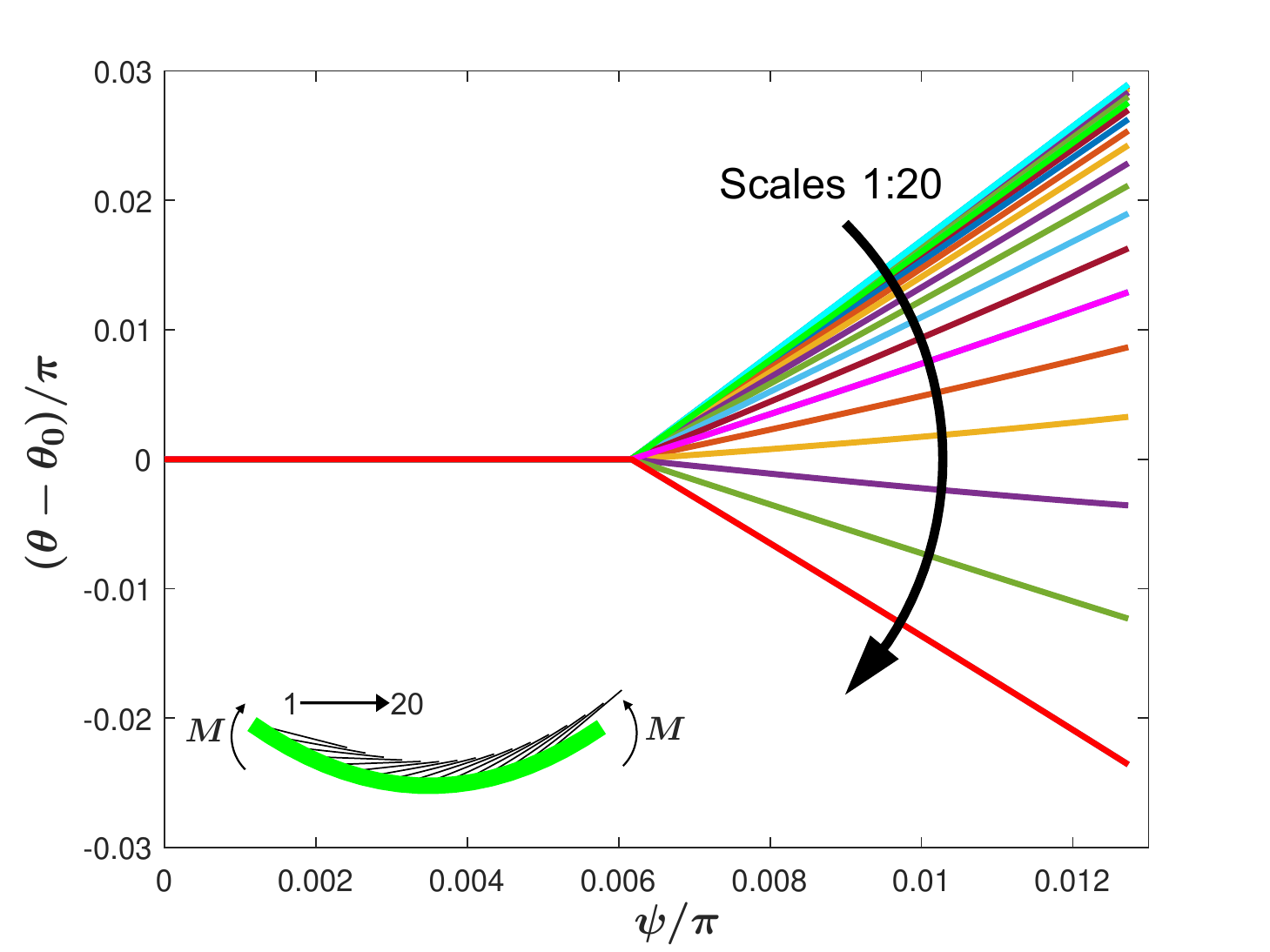}
\label{fig5A}}
\hfill
\subfigure[]{%
\includegraphics[scale = 0.4]{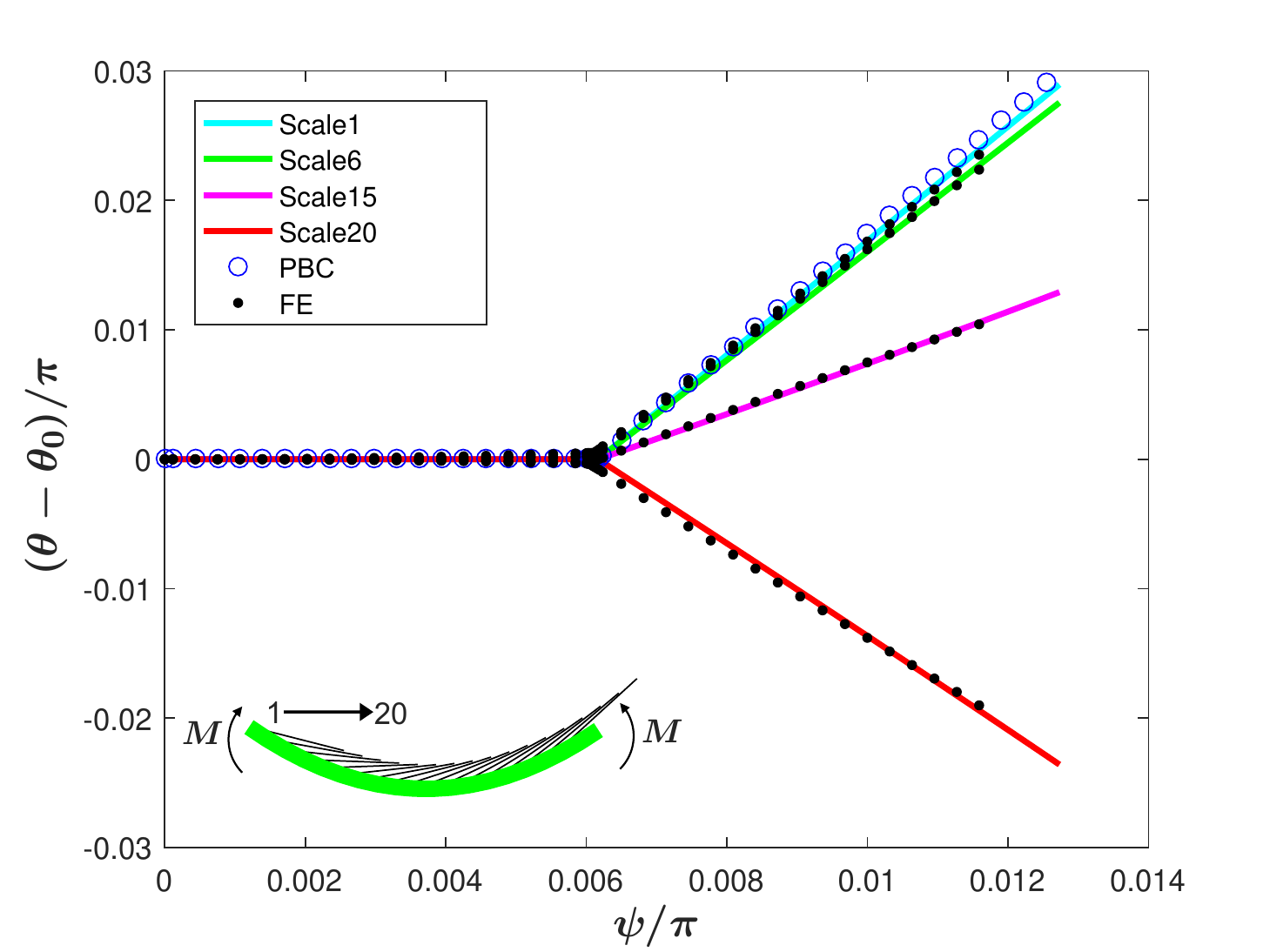}
\label{fig5B}}
\hfill
\subfigure[]{%
\includegraphics[scale = 0.4]{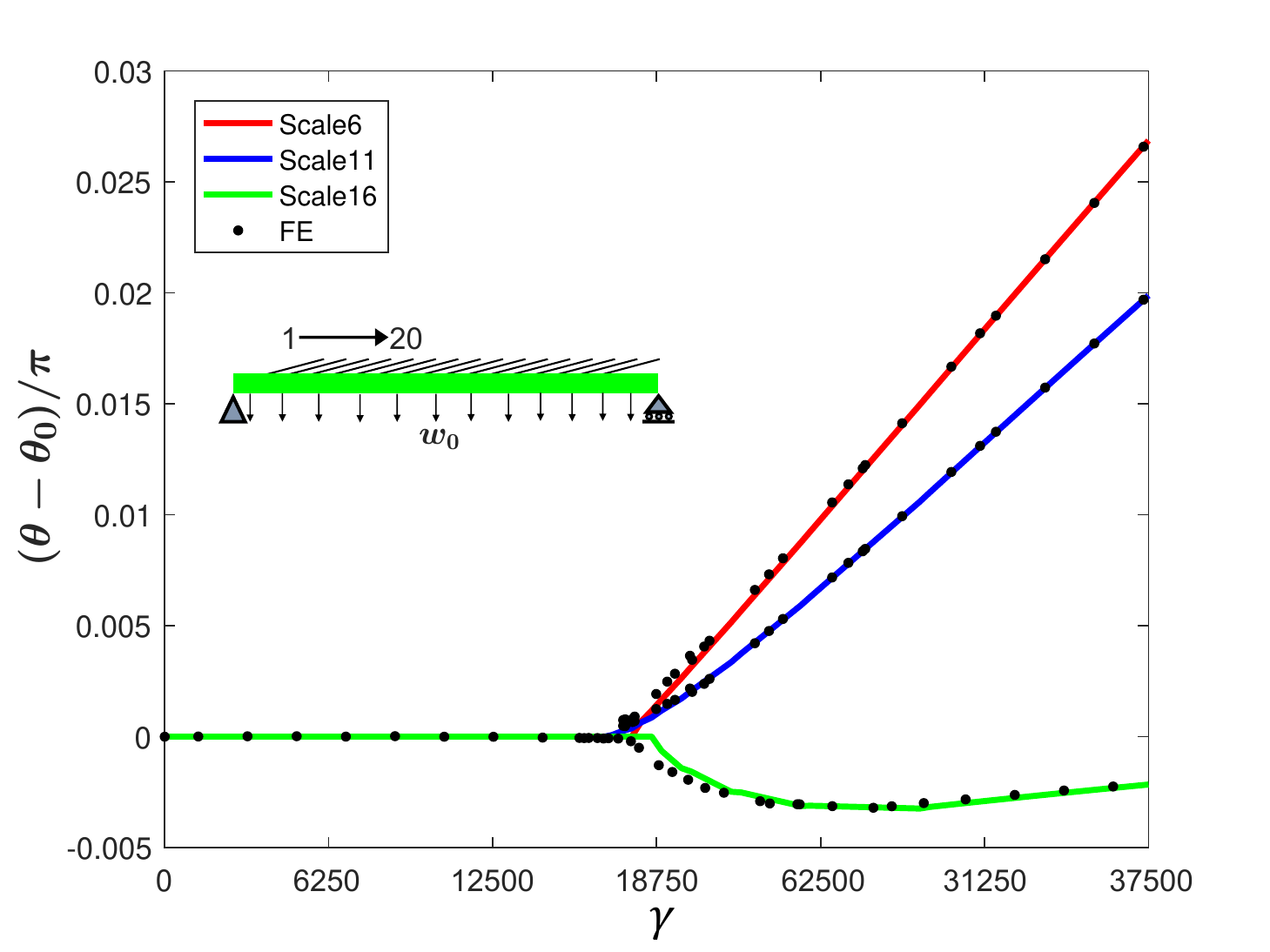}
\label{fig5C}}
\hfill
\subfigure[]{%
\includegraphics[scale = 0.4]{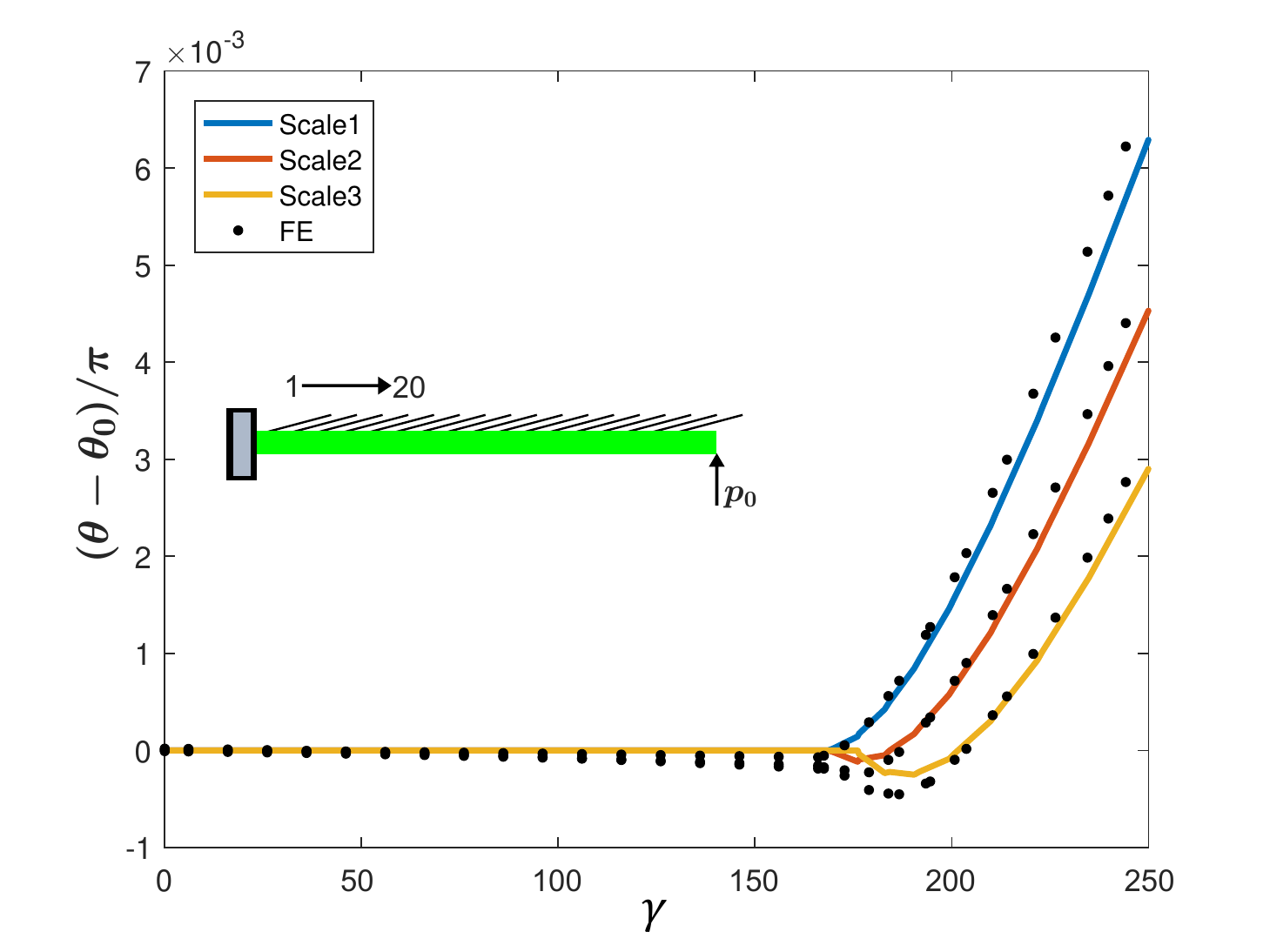}
\label{fig5D}}
\caption{(a) The angular rotation of  $20$ scales embedded on the top layer of a substrate subject to a pure bending moment. (b)  The solid lines depict the inclination angle of four randomly-selected scales from the $20$ scales.  The hollow circles are the analytical solution based on a periodic boundary condition (PBC) that assumes the angular rotation to be the same for all scales ~\cite{mechanics2}, and the black dotted circles illustrate the FE results. (c) The angular displacement of three scales chosen from the same scaly beam with the imposition of simply supported boundary conditions. (d) The change in angle of three scales when the scaly beam was constrained to deflect as a cantilever type beam.}
\label{fig:figure}
\end{figure}

The previously developed analytical formulation of the kinematics can be used to study the scale angles for uniform bending. Such calculations reveal the extent of periodic engagement of scales  by tracking the motion of all the $20$ scales in the structure with a total of $20$ scales with overlap ratio $\eta=5$. The results are illustrated in Fig.~\ref{fig5A} in which the angular displacement of the scales $\theta$ is plotted versus the rotation of the underlying substrate $\psi$. The plots indicate same angles for all scales (horizontal line) until engagement curvature is reached, after which scales begin to change angles due to scale sliding. However, an important distinction arises from previous studies even for this case. Here, the scales angles begin to differ from each other violating periodicity. The scales are numbered $1-20$ starting from the left side as shown in the inset. The scales on the left of the mid-point (scale number $10$) increase in angle as expected from previous periodic theory. However, scales on the right of this point begin to decrease in angle. The verification of these predictions are carried out with FE simulations of an identical system for a few select scales (in this case selected randomly as number $1$,$6$,$15$, and $20$) and depicted in Fig.~\ref{fig5B}. The figure also compares this model with periodicity assumption used in the literature. Clearly, the current model shows an excellent match with the FE simulations for the kinematics.  

Non-periodic engagement of scales was also observed when the underlying substrate undergoes a non-constant curvature deformation. This makes periodicity impossible from the outset requiring using the presented analytical formulation. First, a simply supported beam subject to a uniform loading $w_0$ is studied. In this case, $\gamma ={w_0h^3 \over 24EI}$ since $\gamma$ quantifies the amplitude of the deflection of the beam and serves as proxy to curvature of previous plots. The results shown in Fig.~\ref{fig5C} illustrates scale rotation angles with $\gamma$  for select scales $6$,$11$, and $16$ for brevity. The developed model once again gives excellent match with FE results. Note that the scales angles variation with deformation is not necessarily linear.  Even more interestingly, a symmetry in the loading and boundary conditions did not lead to any symmetrical behavior in the scales kinematics. Clearly, the scale `handedness', i.e. inclination of the scale played a crucial role in this symmetry breaking. Furthermore, scales engagement begins in the positions that possess higher curvature as the substrate continuously deforms. In the simply supported scaly beam, scales start engaging from the middle and then continue outwards from the center of the beam toward the edges. Additionally, the results show that the angle of scales placed near the right edge of the beam reduces until it touches the subsequent scale and thereafter starts increasing. 

The other example presented to study scales angles of non-uniform bending is the deformation of a scaly cantilever type beam with a point load applied at the tip. The scales angles are plotted versus $\gamma={P_0h^2\over6EI}$, and the results are depicted in Fig.~\ref{fig5D}. The figure clearly shows the non-periodicity of scales engagement. The asymmetry in the structure provides an increase in the scales angles after the engagement with a subsequent scale. It is noticeable that a cantilever scaly beam requires a small $\theta_0$  in order for scales to engage early unlike the case of a simply supported scaly beam. Higher $\theta_0$ will require much higher deformation to engage making the substrate stretch, which is neglected in the developed model. This seems to be the reason for the slight deviation in the results when comparing with FE. This could be an important factor for higher angle scales, although they are not typically considered to be as useful due to late engagement. 

\begin{figure}[h!]
\centering
\subfigure[]{%
\includegraphics[scale = 0.6]{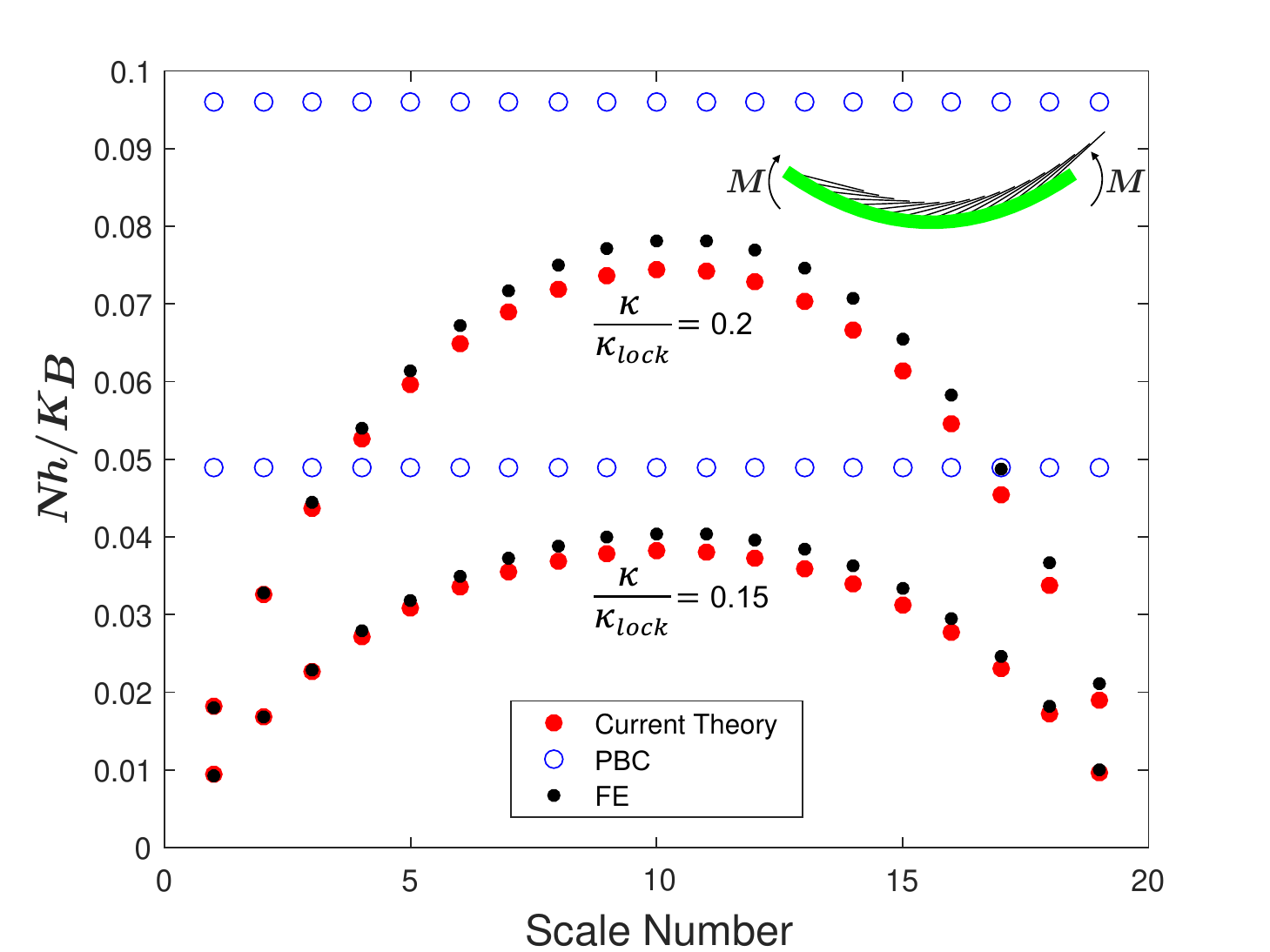}
\label{fig6A}}
\quad
\subfigure[]{%
\includegraphics[scale = 0.4]{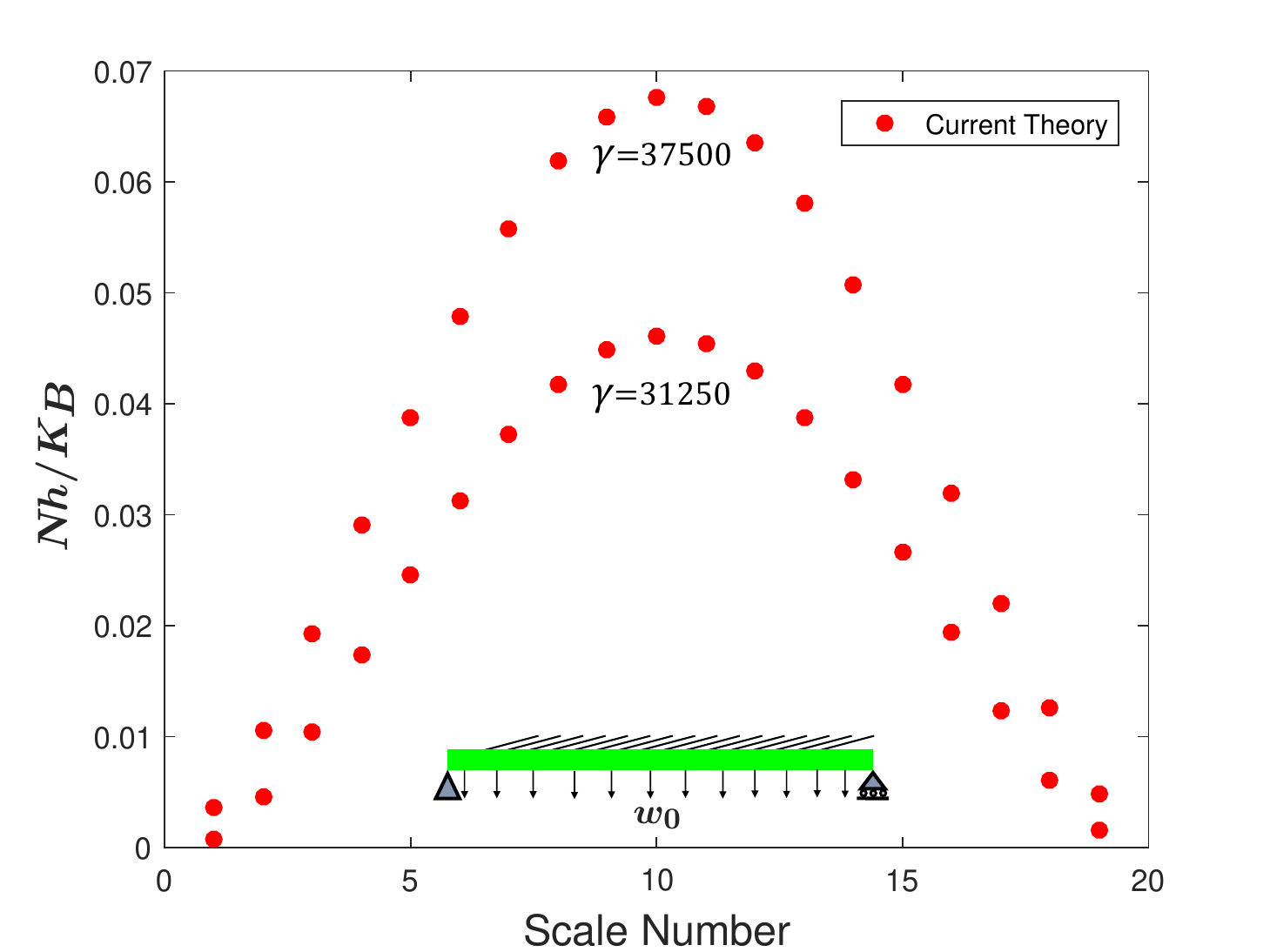}
\label{fig6B}}
\subfigure[]{%
\includegraphics[scale = 0.4]{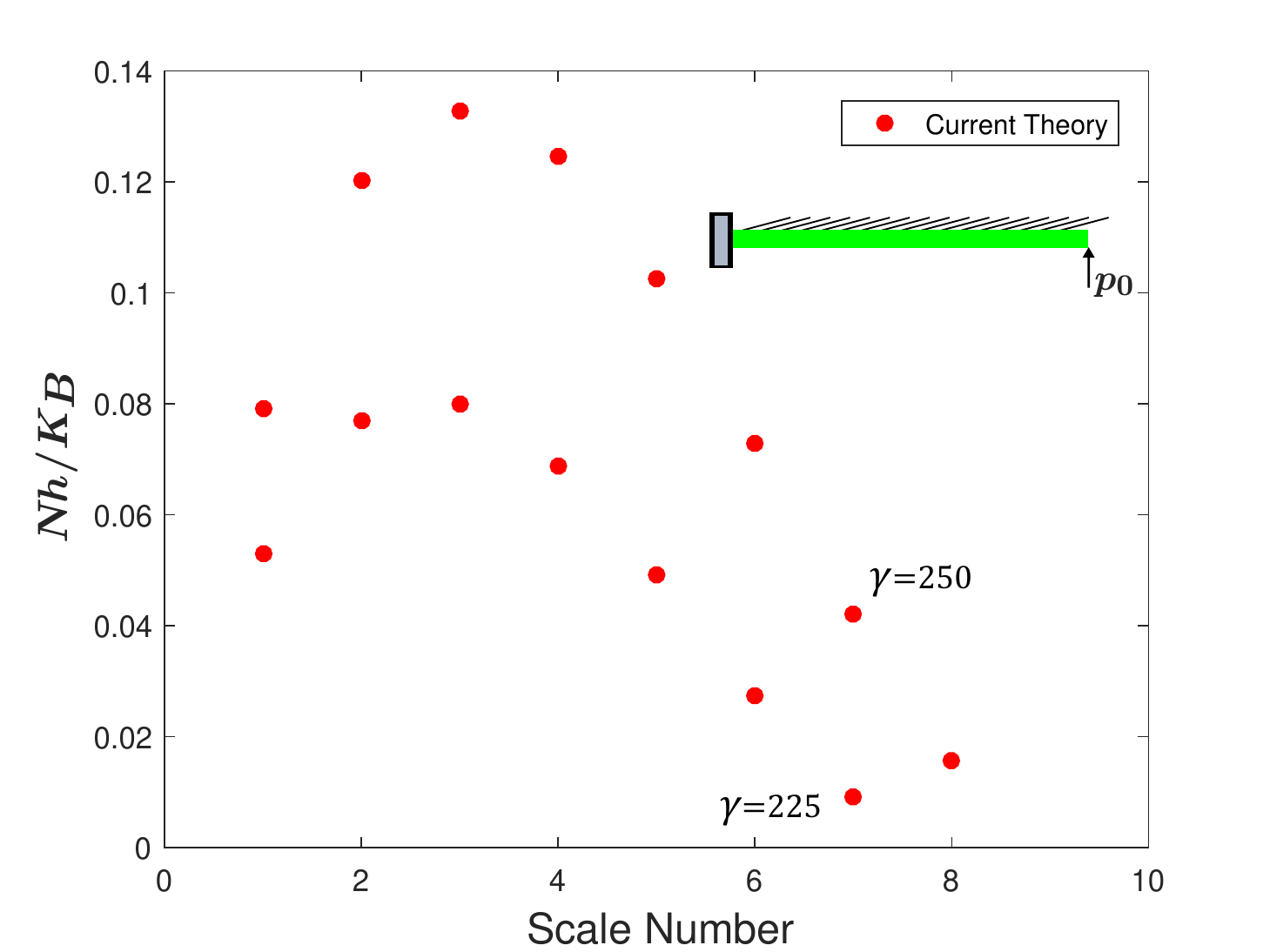}
\label{fig6C}}
\caption{(a) The red dots represent the non-dimensionalized normal force between scales after engagement for two cases of  $\kappa/\kappa_{lock}$   when the beam experiences a uniform bending. The hollow circles are the results of the periodic boundary condition assumption (PBC) ~\cite{mechanics3} and the FE results are shown using the black dots. (b) The variation of the normalized normal force between scales after engagement of a simply supported beam with 20 scales. (c)  Non-dimensional reaction force between scales for the case of a cantilever scaly beam.}
\label{fig:figure}
\end{figure}

These results also forces a re-discussion on the limits of nonlinearity i.e. locking behavior (bending rigidity sharply increases due to transition from substrate to stiff scale bending) at which the sliding of the scales would eventually stop~\cite{mechanics1,mechanics2}. For periodic engagement of rigid scales, a relationship that relates the locking angle of the scale $\theta_{lock}$ to the substrate unit cell rotation $\psi$ was derived earlier $\theta_{lock}+\psi_{lock}/2=\pi/2$~\cite{mechanics2}. This formula was derived based on studying the kinematics of a single RVE due to imposition of periodicity. The periodicity of the geometry makes any further motion geometrically impossible. The normal force (see Fig.~\ref{fig3B} ) at this point is singular and same for all scales. However, in practical cases this point is never reached due to scale deformation or frictional effects even for minor coefficients of frictions~\cite{mechanics3}. In the current problem, the lack of periodicity precludes a kinematic lock. However, considering the critical importance of the normal reaction force, locking could be reformulated on the basis of normal reaction force. The normal force can be determined employing Eqs.~\eqref{eq6} and ~\eqref{eq8} and plotted for all embedded scales in Figs.~\ref{fig6A} through ~\ref{fig6C}. Normal force will not be constant due to lack of periodicity. In fact, calculations in this paper reveals that the normal force which has been previously assumed to be the same for all scales when a scaly structure undergoes a uniform bending is not always true.  The normal force in the results is normalized by the product of height of the beam  $h$ and the spring constant $K_B$. For the case of uniform bending, the theory developed above revealed that the non-dimensional normal force follows a parabolic shape, which indicates that the structure begins locking from the middle of the beam. Fig.~\ref{fig6A}, compares the normalized reaction forces utilizing the developed theory (Eqs.~\eqref{eq6} and ~\eqref{eq8}) and the previous work with FE for the cases of ${\kappa \over \kappa_{lock}}= $ 0.15 and 0.2 for pure bending. The $\kappa_{lock}$ was calculated following the formula $\theta_{lock}+\psi_{lock}/2=\pi/2$. The figure also compares the constant normal reaction arising from the periodicity assumption at any given curvature. However, in reality this is not the case even for pure bending with maximum normal force in the middle which then decreases near the edges as shown in FE simulations, Fig.~\ref{fig6A}. This phenomenon is accurately predicted by the currently developed theory. The periodic theory also over predicts the normal reaction, which is also corrected in this work. However, for periodic contact, an ideal case for locking is a kinematic limit although it is likely that the spike in normal reaction in the middle of the mid prevents locking far earlier than kinematic prediction via deformation or friction (which would no longer remain negligible).

The theory also demonstrates that locking in symmetric scaly structures begins at the middle of the structure, and that is true even for the case of non-uniform bending of a simply supported beam subject to a uniform loading as depicted in Fig.~\ref{fig6B}. The figure illustrates the normalized reaction force between the scales for the two cases of  $\gamma= 31250$ and $37500$. On the contrary, the current results show that non-symmetric scaly beams will start locking near the edge that is exposed to the highest curvature. The results of tracking the force between scales in the cantilever scaly beam is shown in Fig.~\ref{fig6C} for two cases of  ${\gamma}= 225$ and $250$. Finally, the presented theory demonstrated that locking would not take place globally in the structure, but in a more gentle progressive fashion.

\begin{figure}[t!]
\centering
\subfigure[]{%
\includegraphics[scale = 0.6]{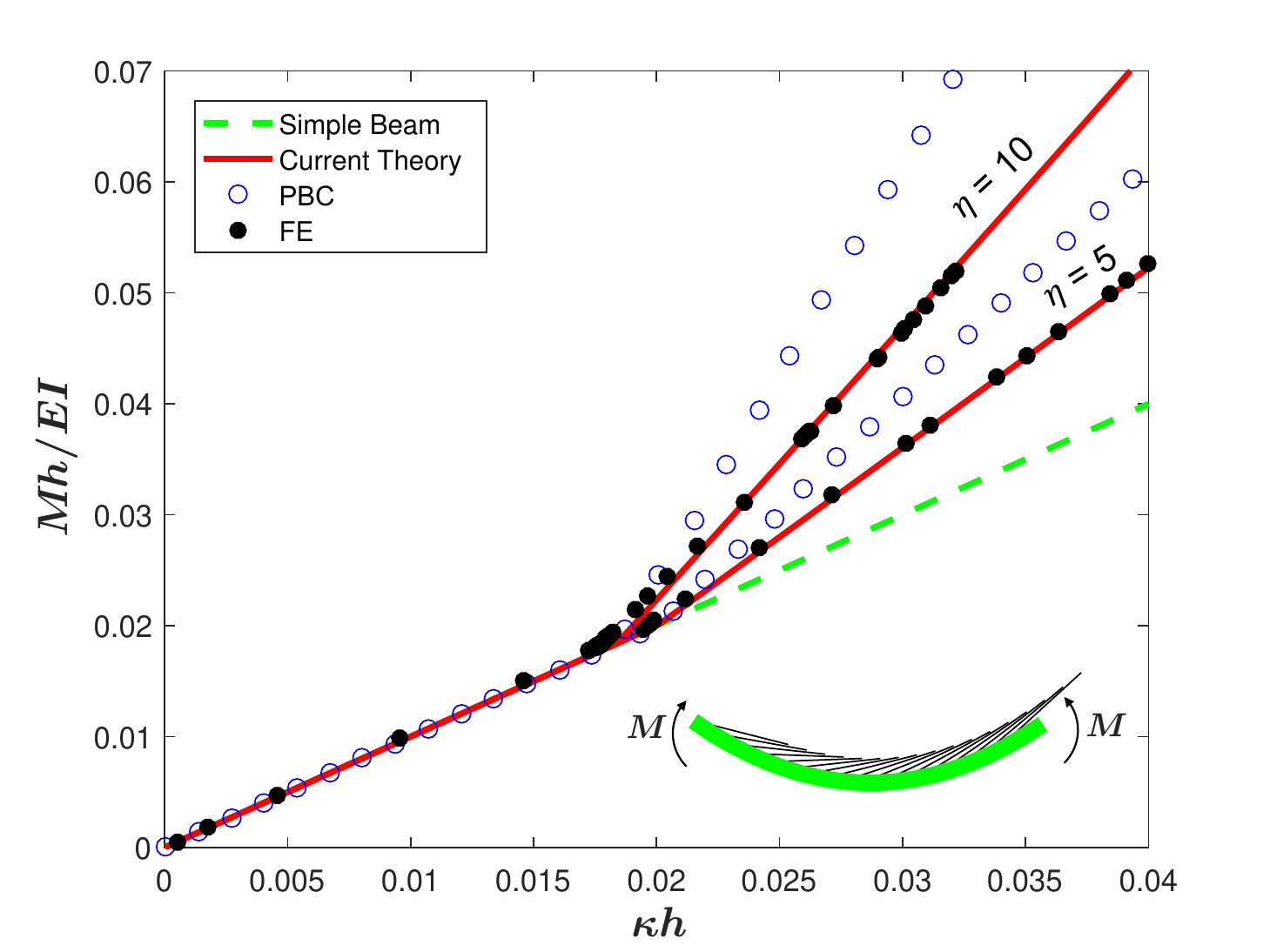}
\label{fig7A}}
\quad
\subfigure[]{%
\includegraphics[scale = 0.4]{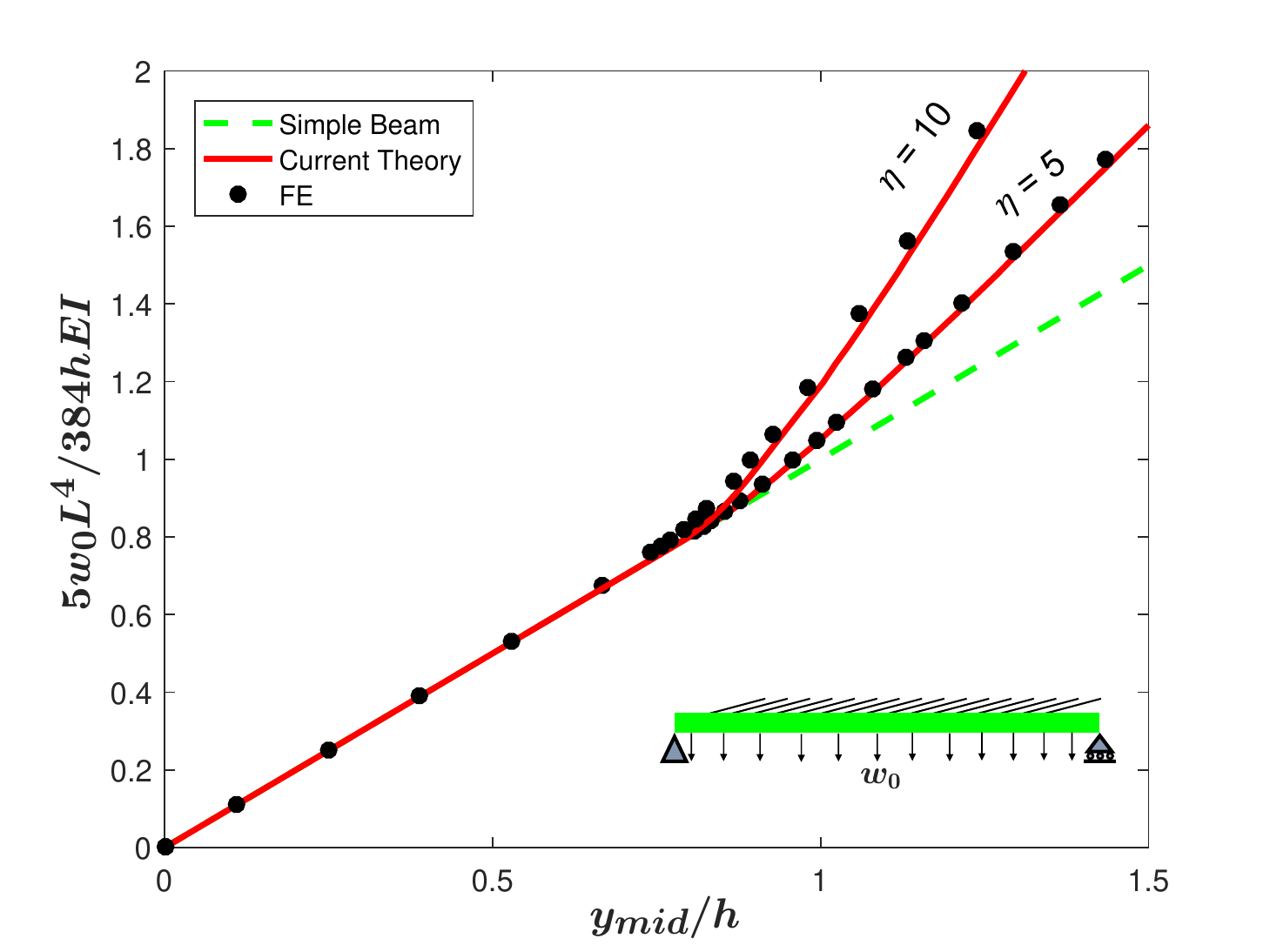}
\label{fig7B}}
\hfill
\subfigure[]{%
\includegraphics[scale = 0.4]{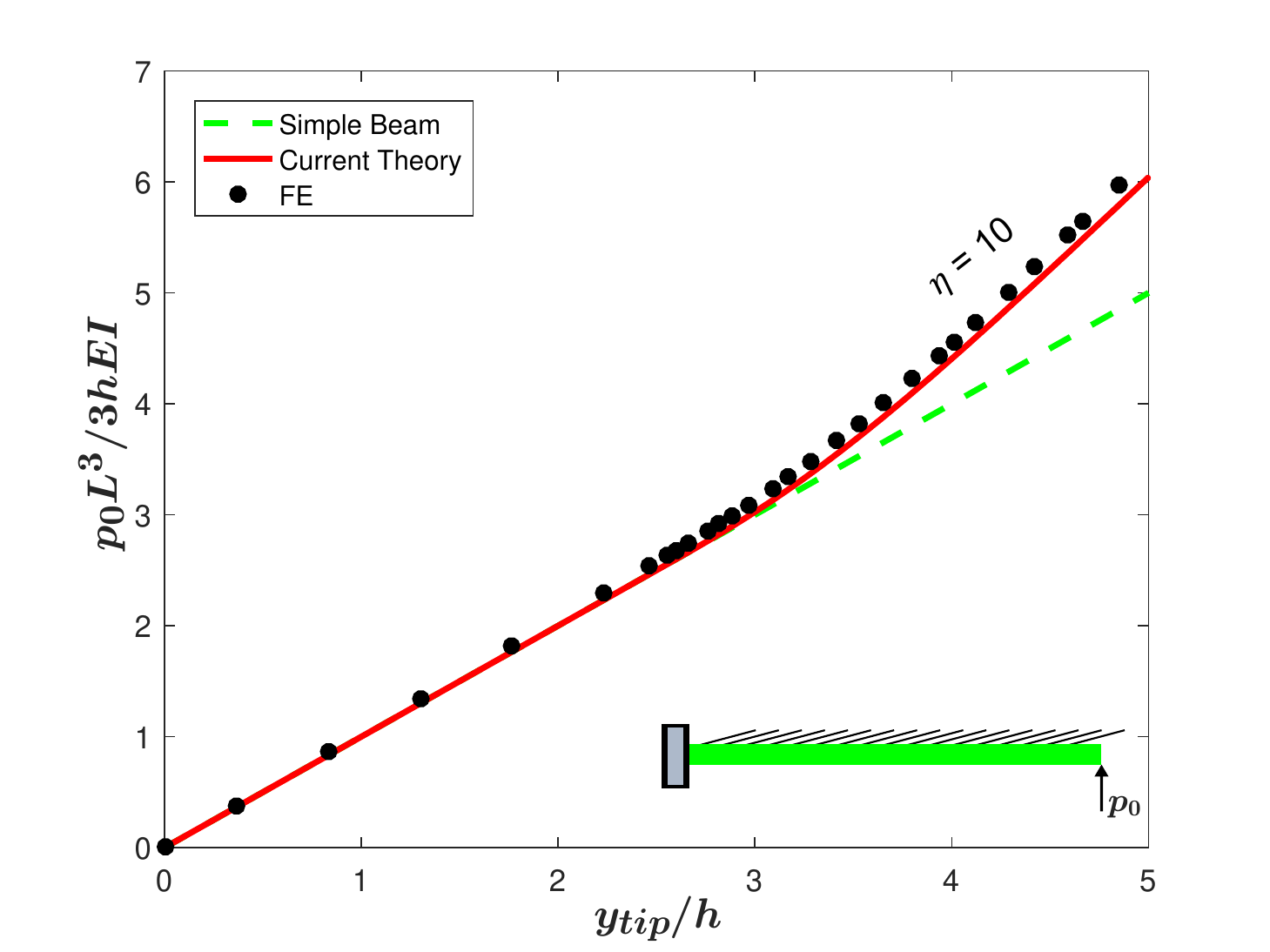}
\label{fig7C}}
\caption{(a) A comparison of the non-dimensionalized moment-curvature relationship of a scaly beam with different $\eta$ using the current method, a PBC: periodic boundary condition assumption previously presented in~\cite{mechanics2}, and FE . (b) The stiffness gained in the deflection of a simply supported scaly beam for different $\eta$ due to scales interaction. (c) The deviation in the tip deflection of a cantilever scaly beam from linearity due to the higher engagement ratio of scales.}
\label{fig:figure}
\end{figure}

Turning now to mechanics to calculate load-displacement like characteristic, the developed model results in an excellent match between our results and those of FE simulations for all these cases.  In the next examples, the results of the mechanical behavior of scaly beams have been normalized by the height of the beam. Figure.~\ref{fig7A} depicts the non-dimensionalized moment-curvature relationship and illustrates how the overlap ratio plays a crucial role in stiffening the structure. The results are plots of the moment curvature for two cases of  $\eta=5$ and $10$. Our current theory exhibits an excellent match with the computational models, correcting previously reported deviations completely. This shows that simply allowing for non-periodicity is sufficient to capture most of the small deformation nonlinear mechanics of these substrates. 

Additionally, the normalized mid-deflection of a simply supported scaly structure was plotted versus the solution obtained from the linear theory of the deflection of beams~\cite{shigleybook}, and the results are shown in Fig.~\ref{fig7B}. Again, an addition in the stiffness of the underlying substrate requires higher $\eta$ , which can be increased by either increasing $l$ or decreasing $d$. Note that lowering $d$ between scales may delay the engagement of scales unlike increasing $l$, a direct conclusion from the vanishing distance parameter. The figure also exhibits a good match with the results obtained from FE.

For the cantilever beam, not much difference was found from the virgin beam for $\eta=5$ in contrast to the simply supported beam. This is because the curvature was not large enough to engage sufficient number of scales. Therefore, for cantilever simulation $\eta=10$ was utilized to effect an appreciable stiffness gain,  Fig.~\ref{fig7C}. It is worth noting that even for the case of this higher $\eta$, not all embedded scales has been engaged due to the low curvature near the tip of the cantilever scaly beam. The results shown in Fig.~\ref{fig7C} show an excellent match between our theoretical model and computational results.


\section{Conclusions}
This paper presents an accurate and validated model for biomimetic scale based system relaxing the previous periodicity assumptions which would not be physical for a realistic system. This is a significant step towards developing the structure-property-architecture framework for topologically leveraged solids such as these opening way to better integration with additive manufacturing and possible topology optimization. The model introduces a new and more accurate way to predict the mechanical properties of the scale covered substrates. The analytical predictions for three test cases have been derived and thoroughly validated with finite element calculations. It was found that non-periodic post engagement behavior cannot be neglected as the errors could be significant. In the same vein, incorporating periodicity eliminated most of the discrepancies of the previous models completely thereby showing no further source of inaccuracies in the previous models.  Using non-periodic general theory allows us to interpret locking more accurately since the original formulation depends on a simultaneous, locked position. It was found that locking in symmetric scaly structures begins at the middle of the structure and continues outward towards the edges. On the other hand, for the case of non-symmetric scaly beams, locking starts near the edge that is exposed to the highest curvature. Symmetric structures require less of an overlapping ratio than non-symmetric structures in order to gain a noticeable stiffness. This is important for a number of applications such as substrate design, soft robotic gripper, deployable structures etc. which would exhibit complex non-periodic and discrete type mechanics.

\section*{References}

\bibliography{myreferences}
 \end{document}